\RequirePackage{fix-cm}
\documentclass[smallextended]{svjour3}       
\smartqed  
\usepackage{graphicx}
\usepackage{gensymb}
\usepackage{amsmath,amssymb}
\usepackage{epsfig,here}
\usepackage{mathptmx} 
\usepackage{epsfig}
\begin{document}

\title{Comments on ``Comment on ``Finiteness of corner vortices" [ Z. Angew. Math. Phys. (2018) 69:37]"[Z. Angew. Math. Phys. (2018) 69:64]}


\author{Jiten C. Kalita}


\institute{
              Jiten C. Kalita (Corresponding Author) \at
              Department of Mathematics\\
Indian Institute of Technology Guwahati\\
PIN 781039, INDIA. \email{jiten@iitg.ernet.in}           
}

\date{Received: date / Accepted: date}

\maketitle

The notion of infiniteness as pointed out in \cite{kal18,sh18} comes from the solution of the biharmonic equation
\begin{equation}\label{bi}
\nabla^4\psi=0
\end{equation}

The above is a simplified version of the Navier-Stokes equations for creeping flow. The author of \cite{sh18} V. Shtern contends that although the solution of \eqref{bi} leads to a sequence of infinite number of eddies, in physical reality, the sequence of eddies is finite. This is a clear admission that \eqref{bi} does not accurately model the physical reality. In a self-contradictory defence, the author (\cite{sh18}) comments that `` ``the sequence of corner eddies is finite", which is valid from the physical point of view, is incorrect for the mathematical solution." 

It is highly probable that tweaked statements such as ``incorrect for mathematical solution" and "valid from physical point of view" could divert the scientific community further from the ground reality, which, on the contrary is not very difficult to fathom in the current context! If a partial differential equation modelling a physical system produces a mathematically correct solution having no physical relevance, it is a clear indication that the equation has failed to  model the physical system correctly. Viscous effects on the wall, which is a pre-requisite for flow separation leading to the formation of vortex is completely absent in equation \eqref{bi}.  We have also reiterated in our original work \cite{kal18} that one of the sources of this non-physical result is the discontinuity of the boundary conditions at $r=0$, ie., at the corner and  failure of equation \eqref{bi} in adhering to the continuum hypothesis on which the foundation of Navier-Stokes equation is built upon. R. L. Panton, in his highly acclaimed book {\bf Incompressible flows} \cite{pan} concurs with our views: citing a specific case of Stokes flow in corners, he states that the remedy of removing this discontinuity lies in solving another problem that allows for a small gap at the corner (the junction of the walls). If the gap is actually smaller than the continuum length scale, then there is a failure of the continuum hypothesis in this boundary condition.  

Shtern \cite{sh18} further claims that ``the solution by Dean and Montagnon \cite{dean}, which was interpreted by Moffatt \cite{moff2}, is not analytical at the corners, therefore the theorems on a finite number of critical points in bounded domain by Kalita {\it et al.} \cite{kal18} are not applicable to the corner problem". This is a completely vague, out of the context and misleading statement, and there is no connection between the notion of infiniteness/ finiteness and a solution being analytic at the corner. Most of the books on elementary differential equations \cite{sne,john,zan,chil} would reveal that a boundary value problem (BVP) in a bounded domain contains two separate set of equations, one valid at the interior requiring it to be analytic or regular, and the other set of equations being prescribed at the boundaries. Equation \eqref{bi} or for that matter the equations under scrutiny in \cite{kal18,bakker,delery,ghil,hir,wang,wu}  are no different from them.

The differential equations in the context of the current discussion pertain to a closed physical domain bounded by solid walls. Analyticity of the solution at the boundaries of such bounded domains (which invariably includes the corner) is not guaranteed because the derivative of a function is not defined at the boundaries. Owing to this, analytic functions are always defined on an open set. 
For the benefit of the readers, we offer a few lines from the book {\bf Complex variables and applications} by Brown and Churchill \cite{bro} on the definition of an analytic function: "A function $f$ of the complex variable $z$ is analytic in an open set if it has derivative at each point in that set. If we should speak of a function $f$ that is analytic in a set $S$ which is not open, it is to be understood that $f$ is anlytic in an open set containing $S$. In particular, $f$ is {\it analytic at a point $z_0$} if it is analytic in a neighbourhood of $z_0$." Opting to go for a complex function above arises out of the fact that the solution of \eqref{bi} in \cite{sh18,dean,moff2} is in the form of a complex variable.

Nowhere in the theorems of \cite{kal18} and the theorems preceding it \cite{bakker,delery,ghil,hir,wang,wu} state the solution of  Navier-Stokes equations to be analytic at the corner. Note that the geometric theorems on incompressible viscous flows \cite{ghil,wang} consider a divergence free vector field as the solution at the interior (which is regular in an open set) of a bounded domain and Dirichlet or Neumann condition at the boundaries. The two main proofs in \cite{kal18} do not use the concept of analytic functions as solutions at all. Moreover, the concepts in \cite{bakker,delery,ghil,hir,wang,wu} which are used to derive the alternative proofs of finiteness in \cite{kal18}, employ regular functions defined in an open subset, and hence are devoid of analyticity requirements at the boundary, which the corner is a part of. The endeavour of connecting the notion of finiteness and a solution being analytic at the corners is therefore completely far fetched.

The very definition of vortex requires \cite{hal} the rotational effect in the immediate neighbourhood of the vortex center. As such the circulation \cite{white,batch} of the vortex must be non zero. On the other hand, a vortex with zero length scale will naturally render a zero circulation for a vortex and hence would fail to satisfy the definition of a vortex. Therefore the smallest length scale that can be accommodated in the so called infinite sequence of vortices must be non-zero, even if smaller than the Kolomogorov length scale. This scale was used in the main proofs on finiteness in \cite{kal18} to signify a non zero length scale only. Any such length scale (referred to as $\eta$ in \cite{kal18}, page 10-11), be it Kolmogorov or anything else, is sufficient to ensure the finiteness.  Thus the claim by the author of reference \cite{sh18} that the notion of finiteness ``is not even heuristically valid if there are geometric length scales which are smaller than the Kolmogorov scale" is an invalid one. 

The above brings us to another point raised in \cite{sh18}, that the authors (of \cite{kal18}) had not offered or referred to any proof of the statement ``A vortex scale cannot be smaller than Kolmogorov length scale". A close examination of \cite{kal18} would reveal that no such statement was made thereat; rather it was stated that ``the Kolmogorov theory \cite{wal} asserts that eddies below a certain size cannot be formed." However, Shtern \cite{sh18} avoids making any comments on the justification provided in the last paragrpah of section 2.3.1  pertaining to Kolmogorov length scale in \cite{kal18}. Under the same geometric configuration, for a flow undergoing laminar to turbulent transition, can the smallest possible laminar scale  be smaller than the smallest turbulent scale defined by the Kolmogorov scale? The answer is a clear no, which comes from experimental and numerical evidences. For example, the visualization of flow past bluff bodies at high Reynolds numbers, where the flow is laminar just behind the body initially and becomes turbulent in the downstream region away from the body  later on (see the visualizations  in \cite{dyke,zad1,zad2}, Chapter 7 of \cite{white} Chapters 5, 6 of \cite{white1}). Another example in this context is the visualization of flows arising out of plane jets at different Reylonds numbers depicting the laminar and turbulent  regimes \cite{sur,tenn}. Just behind the jets, the turbulent length scales are much smaller than the laminar ones here. Recall that the concept of the Kolmogorov length scale comes from the cascading effect where energy is transferred from larger to smaller eddies. Laminar eddies are not small enough to accommodate the mixing of the eddies to facilitate the transfer of energy \cite{white1}.




%

\end{document}